\documentclass[aps,prb,preprint]{revtex4}

\usepackage{graphicx}

\usepackage{amsmath}

\begin{document}

\title{On the $c$ equivalence principle and its relation to the weak equivalence principle of general relativity}

\author{T. C. Choy}
\email{tuckvk3cca@gmail.com}
\affiliation{London Centre for Nanotechnology and Department of Physics and Astronomy, University College London, Gower Street, London WC1E 6BT, UK}

\begin{abstract}

We clarify the status of the $c$ equivalence principle ($c_u=c$) recently proposed by Heras et al \cite{JoseAJP2010,JoseEJP2010} and show that its proposal leads to an extension of the current framework of classical relativistic electrodynamics (CRE). This is because in the MLT (mass, length and time) system of units,  CRE theory can contain only one fundamental constant of nature and special relativity dictates that this must be $c$, the standard speed of light in vacuum, a point not sufficiently emphasized in most textbooks with the exception of a few such as Panofsky and Phillips \cite{PanofskyPhillips}.  The $c$ equivalence principle Heras \cite{JoseAJP2010,JoseEJP2010} can be shown to be linked to the second postulate of special relativity which extends the constancy of the unique velocity of light to all of physics (especially to mechanics) other than electromagnetism. An interesting corollary is that both the weak equivalence principle of general relativity and the $c$ equivalence principle are in fact one and the same, which we demonstrate within the context of Newtonian gravity.

\end{abstract}

\maketitle

\section{Introduction}
\label{Introduction}
In recent papers Heras \cite{JoseAJP2010,JoseEJP2010} suggested a $c$ equivalence principle to identify the equality of the $c_u$ as obtained from static measurements for action-at-a-distance laws such as Coulomb's (his eqn(5)) with the speed of propagation of light $c$ \cite{Note1, Note2}. This principle was proposed as a way to ``explain" the numerical coincidence of both $c_u$ and $c$ in a way analogous to the mass equivalence principle for the gravitational and inertial mass, which is a key ingredient for the success of general relativity.  The question that naturally arises is: what is the status of this principle?  Although not stated in his AJP publication \cite{JoseAJP2010} but elsewhere Heras \cite{JoseEJP2010} and his collaborators \cite{HerasBaezEJP2009} seem to claim that this is a new principle which should be invoked as an {\it additional} postulate that cannot be derived from current theories.  On the other hand, most readers would be puzzled by this proposal, if in fact something is adrift in our current theory of classical relativistic electrodynamics (CRE) and that the proposed $c$ equivalence principle of Heras et al \cite{JoseAJP2010,JoseEJP2010} while novel, is not fundamental and can thus be derived from special relativity or elsewhere if we accept in particular Einstein's second postulate of special relativity as universal. In this paper, I shall clarify some of these issues. After that I shall apply our results to the study of the weak equivalence principle, widely accepted to be fundamental within the context of general relativity but there are speculations that it can be derived from a deeper theory such as quantum gravity \cite{Weinberg}.  An interesting corollary is that both the weak equivalence and and the $c$ equivalence principle are one and the same, within the context of Newtonian gravity.

\section{Gauge Invariance and Special Relativity}
\label{Gauge Invariance and Special Relativity}
We shall first show that the $c$ equivalence principle is outside the frame work of CRE if we take into account the Minkowski spacetime structure in which $c$ is the appropriate universal constant for the time component of spacetime coordinate four vectors \cite{Landau1}, such as $x^i = (ct,\bf{x})$ {\it and} also the principle of gauge invariance of electrodynamics \cite{Choy1}.
To see this let us look at ``the correct form of Maxwell's equations" as proposed by Heras et al \cite{JoseEJP2010} in their so called $\alpha, \beta, \gamma$ units:
\begin{eqnarray}
{\bf \nabla . E} = \alpha \rho, \nonumber \\
{\bf \nabla . B} = 0, \nonumber \\
{\bf \nabla \times E}+ \gamma {\partial {\bf B} \over \partial t} = 0,\nonumber \\
{\bf \nabla \times B}- {\beta \over \alpha} {\partial {\bf E} \over \partial t} = \beta {\bf J}.
\label{eqn1a}
\end{eqnarray}
Here $\gamma = \chi{c_u^2\over c^2}$ and the various constants $\alpha,\beta$ and $\gamma$ are added to allow for an arbitrary choice of units according to Heras et al \cite{JoseEJP2010}.  The equations of Maxwell are required to satisfy gauge invariance.  This is indeed the case if we were to define the gauge transformations as follows:
\begin{eqnarray}
{\bf A}' = {\bf  A} + {\bf \nabla} \Lambda, \nonumber \\
\phi' = \phi - \chi{c_u^2\over c^2} {\partial \Lambda \over \partial t},
\label{eqn1b}
\end{eqnarray}
with the usual definitions in terms of the vector ${\bf A}$ and scalar $\phi$ potentials respectively.
\begin{eqnarray}
{\bf B}={\bf \nabla \times A} , \nonumber \\
{\bf E} = -{\bf \nabla} \Lambda- \chi{c_u^2\over c^2} {\partial {\bf A} \over \partial t}.
\label{eqn1c}
\end{eqnarray}
However none of the gauge transformation equations (\ref{eqn1b}) are in accord with special relativity.  The latter requires that in Gaussian or Heaviside Lorentz units (where $\chi={1\over c_u}$) that:
\begin{eqnarray}
{\bf A}' = {\bf  A} + {\bf \nabla} \Lambda, \nonumber \\
\phi' = \phi - {1\over c} {\partial \Lambda \over \partial t},
\label{eqn2a}
\end{eqnarray}
or in SI units (where $\chi=1$) that:
\begin{eqnarray}
{\bf A}' = {\bf  A} + {\bf \nabla} \Lambda, \nonumber \\
\phi' = \phi - {\partial \Lambda \over \partial t},
\label{eqn2b}
\end{eqnarray}
unless of course if $c_u=c $.  This shows that the $c$ equivalence principle is intimately connected with {\it both} gauge invariance and special relativity and cannot be deemed to be an additional independent postulate.  Clearly without the $c$ equivalence principle, we are operating outside the regime of special relativity in some way and the obvious question is in what way does this free parameter $c_u \ne c$ do so. We shall answer this question shortly but it is worth reminding the reader on the case of arbitrary units in CRE.

\section{Arbitrary Units}
\label{Aribitrary Units}
The case of arbitrary units in CRE has been expounded in some detail in traditional texts see for example notably Landau and Lifshitz \cite{Landau1}.  It is perhaps worth reiterating that within the frame work of CRE with arbitrary units, there can be only one arbitrary constant denoted by $a$ in Landau and Lifshitz \cite{Landau2} whose action takes the form:
\begin{equation}
S= -\sum  \int mc \  ds - {1 \over c^2} \int  A_i J^i d\Omega - {a \over c} \int  F_{jk}F^{jk} d\Omega,
\label{eqn3}
\end{equation}
where the four vector $J^i=(c\rho,{\bf j})$.

If, as is traditional, we adopt the MLT system (where units of mass, length and time are afforded dimensions) for the mechanical units, then the first term will define the energy and therefore the action to be in such units (e.g. ergs or joules). Then the choice of $a$ automatically sets the units for {\it all} electromagnetic quantities.  It is convenient for special relativity to choose $a={1\over 16 \pi}$ as a dimensionless constant (which defines the Gaussian units) for this then removes any dimensionality associated with $a$ and defines the fields $F^{ik}$ and therefore $A_i$ to share the same dimensions as the mechanical units.  In this way, the current $J^i$ will also acquire mechanical units given by the second term (assuming minimal coupling electrodynamics), in which the unit of current (the abampere) has dimensions $(M^{1/2}L^{1/2}T^{-1})$. This clearly demonstrates as stressed by Panofsky and Phillips \cite{PanofskyPhillips} that the speed of light $c$ is the only fundamental constant of CRE \cite{Note3}.
In the alternative SI units, universally adopted because of engineering convenience, $a$ is given by the choice: $a={\epsilon_0 \over 4}$ or $a={1\over 4\mu_0 c^2}$ which has the electrical units dimensions of farads m$^{-1}$.  Without going through the details \cite{Landau3} we can derive the covariant form for the second pair of Maxwell's equations by the principle of least action upon arbitrary variation of the fields $A_i$. These are given by:
\begin{equation}
{\partial F^{ik} \over \partial x^k}=-{1\over 4 ac}J^i ,
\label{covariant Maxwell}
\end{equation}
while the first pair of Maxwell's equations follows trivially from the antisymmetry of the field $F_{ik}=(\partial_i A_k - \partial_k A_i)$ and does not involve the constant $a$, which we reiterate is the only arbitrary constant for defining the electromagnetic units in the MLT system within the framework of CRE.

\section{$\alpha,\beta,\gamma$ Units}
\label{ABC Units}
We now rewrite the action in the $\alpha,\beta,\gamma$ units of Heras eta al \cite{JoseAJP2010,JoseEJP2010,HerasBaezEJP2009}, which they have given in the form:
\begin{equation}
S= -\sum  \int mc \  ds - {\gamma \over c} \int  A_i J^i d\Omega - {\gamma \over 4 \beta c} \int  F_{jk}F^{jk} d\Omega.
\label{ABCeqn}
\end{equation}
where $\gamma=\chi {c_u^2 \over c^2}$, without invoking the $c$ equivalence principle.  Note that $\gamma$ is a factor multiplying both terms in the last two terms, thus it affects only a trivial rescaling of the energy units and will have no effect on the covariant Maxwell's equations under variation of the fields.  The only relevant parameter is $\beta$ in accordance to the variational principle and setting $\gamma={1 \over c}$ and $a={1\over 4\beta c}$ takes us back to the form eqn(\ref{covariant Maxwell}).  However the choice of $\gamma$ (and the other parameters) does affect the units (and the dimensions) of the electromagnetic energy relative to the mechanical energy.  Hence the choice of the $\alpha,\beta,\gamma$ units without invoking the $c$ equivalence principle amounts to a redefinition of the mechanical units.  This can also be seen by the form of eqn(6) in Heras \cite{JoseAJP2010} which we shall write as:
\begin{equation}
{1\over \chi}{d F \over d\ell}={\beta\over 4 \pi}{2 I^2\over R},
\label{eqn6Heras}
\end{equation}
which allows for a redefinition of the unit (and dimension) for the force or energy, otherwise $\chi$ can serve no useful purpose and we could have redefined $\beta'=\beta\chi$ and eliminate $\chi$ altogether.  Let us now agree to measure energy in units of $\gamma$, (i.e. $S=\gamma\tilde S$) without invoking the $c$ equivalence principle, then equation (\ref{ABCeqn}) now formally looks like equation (\ref{eqn3}) and becomes:
\begin{equation}
\tilde S= -\sum  \int {\tilde m} c \  ds - {1\over c} \int  A_i J^i d\Omega - {1\over 4 \beta c} \int  F_{jk}F^{jk} d\Omega,
\label{ABCeqn2}
\end{equation}
where $\tilde m={m \over \gamma}$ and the rescaling does not affect Maxwell's equations in any way or the particles' trajectory which is still given by ${du^i\over ds}=0$. The particle relativistic Lagrangian is now:
\begin{equation}
\tilde L= - \tilde m c^2 \sqrt{1-{\upsilon^2 \over c^2}}.
\label{Ltildeeqn}
\end{equation}
However unlike CRE here $\tilde m$ is not a free parameter because $\gamma$ and $\beta$ are related to each other. This is a completely different description of nature that is outside current classical electrodynamics (CRE) and the hypothesis now takes the form of a suggested electromagnetic renormalisation of the bare electron mass since  $\tilde m$ is coupled to the electromagnetic field via $\gamma$ which is related to $\beta$.  In particular in the limit  ($c\rightarrow \infty $) from equation(\ref{Ltildeeqn}) and equation(\ref{ABCeqn2}), we have:
\begin{equation}
\tilde {\cal E}= \tilde m c^2 + {\tilde m \upsilon^2 \over 2} + \tilde {\cal E}_{em},
\label{Newtonian energy}
\end{equation}
where $\tilde {\cal E}_{em}$ is the electromagnetic energy in the Galilean limit \cite{JoseAJP2010,JoseEJP2010} but the rest energy and the kinetic energy terms are now dependent on the electromagnetic constant $\beta$ or inherently $c_u$ unless $c_u=c$. Since $\tilde m$ is now of order $O(c^2)$, ( we assume $c_u<c$ on the grounds that $c$ must be the maximum interaction velocity \cite{Landau1}), we can now see that the choice of $c_u \ne c$ violates special relativity as it implies the rest energy now scales as $c^4$ while the kinetic energy scales as $c^2$ in the $\gamma$ description.  As physics must be invariant to scale transformations (i.e. units can be chosen arbitrarily), such a description is unacceptable. In a way here we can see that $c_u=c$ is in fact a restatement of Einstein's'second postulate of special relativity which extends the universal constant $c$ to all physical phenomena and not just electromagnetic ones.  In the words of Panofsky and Phillips \cite{PanofskyPhillips}, ``Physical laws thus ``scale" correctly over arbitrary magnitudes only if the ratios of length and time are held constant"  ( by the universality of the constant $c$).   Of course special relativity does not hold in the presence of the gravitational field, except locally.  So what can we learn by not invoking the the $c$ equivalence principle in this case?  Unfortunately a proper study of this issue will involve general relativity and take us too far afield, so we shall be content with a demonstration within the ``flat spacetime" regime of weak gravity which is rather instructive.
\section{Mass equivalence principle}
\label{mass equivalence principle}
We now include the gravitational field in the toy model in which $c_u \ne c$ and consider only the case of Newtonian scalar gravity for simplicity. Now the action becomes (in the $\gamma$ description):
\begin{equation}
\tilde S= -\sum  \int {\tilde m_i} c \  ds - {1\over c} \int  A_i J^i d\Omega - {1\over 4 \beta c} \int  F_{jk}F^{jk} d\Omega - \sum  \int {\tilde m_g}{\Psi \over c} \  ds - \int {1\over {8\pi \bar \kappa c}} (\nabla \Psi)^2\  d\Omega,
\label{Gravitation1}
\end{equation}
where $\Psi$ here is the Newtonian gravitational potential (in MLT units) and $\bar \kappa=\gamma \kappa$ is the gravitational constant in the $\gamma$ units. However this description is no longer unique.  We can instead also choose:
\begin{equation}
\tilde S= -\sum  \int {\tilde m_i} c \  ds - {1\over c} \int  A_i J^i d\Omega - {1\over 4 \beta c} \int  F_{jk}F^{jk} d\Omega - \sum  \int { m_g}{\tilde\Psi \over c} \  ds - \int {1\over {8\pi \tilde \kappa c}} (\nabla \tilde \Psi)^2\  d\Omega,
\label{Gravitation2}
\end{equation}
where $\tilde \Psi={1\over \gamma} \Psi$ here is now the (renormalized) Newtonian gravitational potential  and $\tilde \kappa={\kappa\over\gamma}$, in which both are now in the $\gamma$ units.  Either way, our toy model now provides an electromagnetic coupling to gravity via the parameter $\gamma$ as in the previous case. We shall not proceed with a variation with respect to $\Psi$ or $\tilde\Psi$ to obtain the gravitational field equation \cite{Landau4}, but note that in the description equation(\ref{Gravitation1}), the mass equivalence principle (${\tilde m_i}={\tilde m_g}$) is trivial and teaches us nothing new \cite{Note4}. However in the description equation(\ref{Gravitation2}) , the mass equivalence principle is (${\tilde m_i}={m_g}$) which teaches us something new. If physical theory must be scale invariant, then all two descriptions must be identical and we can conclude that the mass equivalence principle and the $c$ equivalence principle are in fact equivalent ( with no punt intended)! This can only be possible if all masses are decoupled from the electromagnetic field i.e. ${\tilde m_i}={\tilde m_g}={m_g}$ which implies the c equivalence principle with $\gamma=1$ in SI units, or alternatively we must return to the form equation(\ref{eqn3}) which requires $\gamma={1 \over c}$ for all units in equation(\ref{ABCeqn}), with $\beta$ or $a$ as the only free parameter.

\section{Conclusion}
\label{Conclusion}
In conclusion, we have re-examined the recently proposed $c$ equivalence principle of Heras et al \cite{JoseAJP2010,JoseEJP2010} and our study shows that it does not constitute a new {\it additional} hypothesis within the frame work of classical relativistic electrodynamics (CRE).  Its proposal constitutes a model that is outside the regime of CRE and is equivalent to the proposal of a new system of mechanical units for which energy is measured in $\gamma$ units.  In addition the model leads to a non-trivial mass renormalization that is coupled to the electromagnetic field which is incompatible with special relativity.  By extending the model to include gravity, which special relativity violates, we have shown an interesting corollary that the mass equivalence principle of general relativity also implies the $c$ equivalence principle, if physics remains invariant to scale changes. Nevertheless the extended model may be useful for studying toy models for the electromagnetic origin of mass \cite{Jennison1977} although current theories are in favour of the Higgs mechanism which involves the weak interactions \cite{Berstein2011}. Some further relations with recent studies on the mass energy equivalence might also be of interests \cite{Hecht2011}.
\section{Acknolwedgement}
My thanks to Denise Ottley at UCL for her help in the preparation of this manuscript.


\begin{thebibliography}{10}
\bibitem{Note2a}  If we do not invoke the $c_u=c$ principle, then there is the issue of considering which limit $c_u \rightarrow \infty$ or $c \rightarrow \infty$ and in which order we are considering, an issued not addressed by the previous authors \cite{JoseAJP2010,JoseEJP2010,HerasBaezEJP2009}.
\bibitem{Choy1} I thank a referee for pointing this out to me as my initial proof was based on the use of Gaussian units only, although such a proof should of course not be dependent on the choice of units.
\end{thebibliography}
\end{document}